\documentclass[prd,twocolumn,eqsecnum,amsfonts,amssymb]{revtex4}

\usepackage{graphicx}

\usepackage{bm}

\newcommand{\beq}{\begin{equation}}
\newcommand{\eeq}{\end{equation}}
\newcommand{\beqs}{\begin{eqnarray}}
\newcommand{\eeqs}{\end{eqnarray}}

\begin{document}

\title{Renormalization-Group Behavior of $\phi^3$ Theories in $d=6$ Dimensions}

\author{John A. Gracey$^a$, Thomas A. Ryttov$^b$, and Robert Shrock$^c$}

\affiliation{(a) \ Theoretical Physics Division \\ Department of Mathematical Sciences \\
University of Liverpool, Liverpool, L69 3BX, UK }

\affiliation{(b) \ CP$^3$-Origins \\
Southern Denmark University, Campusvej 55, Odense, Denmark}

\affiliation{(c) \ C. N. Yang Institute for Theoretical Physics and
Department of Physics and Astronomy, \\
Stony Brook University, Stony Brook, New York 11794, USA }

\begin{abstract}

  We investigate possible renormalization-group fixed points at
  nonzero coupling in $\phi^3$ theories in six spacetime dimensions,
  using beta functions calculated to the four-loop level.  We analyze
  three theories of this type, with (a) a one-component scalar, (b) a
  scalar transforming as the fundamental representation of a global
  ${\rm SU}(N)$ symmetry group, and (c) a scalar transforming as a
  bi-adjoint representation of a global ${\rm SU}(N) \otimes {\rm
    SU}(N)$ symmetry. We do not find robust evidence for such fixed
  points in theories (a) or (b). Theory (c) has the special feature
  that the one-loop term in the beta function is zero; implications of
  this are discussed.

\end{abstract}

\maketitle


\section{Introduction}
\label{intro}

A topic of fundamental importance in quantum field theory is the
renormalization-group (RG) behavior of a scalar field theory in $d$
spacetime dimensions.  Here we investigate RG behavior and possible RG
fixed points of three scalar field theories with cubic scalar
self-interactions in $d=6$ spacetime dimensions.  These are denoted
generically as $\phi^3_6$ theories and are defined by the path
integral
\beq
Z = \int \prod_x [d {\phi}(x)] \, e^{iS} \ ,
\label{z}
\eeq
where $S=\int d^6x \, {\cal L}$ and ${\cal L}$ is the Lagrangian density.
For the theory with a (real) one-component scalar, 
\beq
{\cal L}_1 = \frac{1}{2}(\partial_\mu \phi) (\partial^\mu \phi )
- \frac{1}{2} m_0^2 \phi^2 - \frac{g}{3!} \phi^3 \ .
\label{lag1}
\eeq
For the theory with a (complex) scalar field $\phi^i$, $i=1,...,N$,
transforming according to the fundamental representation of a global
SU($N$) theory,
\beq
{\cal L}_2 = (\partial_\mu \phi)^\dagger (\partial^\mu \phi )
-m_0^2 \phi^\dagger \phi - \frac{g}{3!} d_{ijk}(\phi^i \phi^j \phi^k + h.c.) \ ,
\label{lag2}
\eeq
where $d_{ijk}$ is the totally symmetric rank-3 tensor for SU($N$),
and sums over repeated indices are understood.  Since $d_{ijk}=0$ for
SU(2), $N$ is restricted to the range $N \ge 3$ for this theory.  We
will discuss the theory with a bi-adjoint scalar below. The $\phi^3_6$
theories are renormalizable, with a dimensionless coupling, $g$.
These theories are invariant under the redefinition (suppressing
possible indices on $\phi$)
\beq
\phi \to -\phi, \quad \quad g \to -g . 
\label{gsignrel}
\eeq
Because of this invariance, one can, without loss of generality, take
$g$ to be non-negative, and we shall do so henceforth.  For technical
simplicity, we also take $m_0=0$. As is well-known, because of the
cubic scalar self-interaction, the energy of the theory is not bounded
below.  Nevertheless, cubic scalar theories have long been used to provide 
simple examples of perturbative calculations in quantum field
theory. They have also been used in statistical mechanics to model
the Yang-Lee edge singularity \cite{fisher} and percolation \cite{wallace}.
A recent general analysis is \cite{gracey2015}. 
The application to statistical mechanics naturally makes use of a
$d=6-\epsilon$ expansion to obtain estimates of critical exponents;
see also \cite{giombi_klebanov}. Here we restrict ourselves to $d=6$. 
Recently, $\phi^3_6$ theories were used for a test of the $a$
theorem \cite{grinstein_challenge}.

Quantum loop corrections lead to a dependence of the physical coupling
$g = g(\mu)$ on the Euclidean energy/momentum scale $\mu$ at which
this coupling is measured.  The dependence of $g(\mu)$ on $\mu$ is
described by the RG beta function of the theory,
\beq
\beta_g = \frac{dg}{d\ln\mu} \ .  
\label{beta}
\eeq
Because the $n$-loop integrals involve $n$'th powers of the quantity
\beq
\frac{g^2 S_6}{(2\pi)^6} = \frac{g^2}{2^6\pi^3}
\label{intfactor}
\eeq
where $S_d = 2\pi^{d/2}/\Gamma(d/2)$ is
the area of the unit sphere $\| x \|=1$ for a vector $x \in {\mathbb R}^d$, it
is convenient to define the variable
\beq
\bar g \equiv \frac{g}{8\pi^{3/2}}
\label{gbar}
\eeq
and the corresponding beta function $\beta_{\bar g} = d\bar g/d\ln\mu$. This 
beta function has the series expansion
\beq
\beta_{\bar g} = \bar g  \sum_{n=1}^\infty b_n \, a^n \ , 
\label{betaseries}
\eeq
where 
\beq
a \equiv (\bar g)^2
\label{adef}
\eeq
and $b_n$ is the $n$-loop coefficient.  The $n$-loop ($n\ell$)
approximation to $\beta_{\bar g}$, denoted $\beta_{\bar g,n\ell}$, is
obtained by replacing $n=\infty$ by finite $n$ in the summand in
Eq. (\ref{betaseries}).  Because of the prefactor $\bar g = \sqrt{a}$ in
Eq. (\ref{betaseries}), the beta function $\beta_{\bar g}$ always
vanishes at the origin in coupling-constant space, $a=0$. Physically,
this just means that in a free theory, there is no running
coupling since the coupling is zero. The one-loop and two-loop coefficients in
Eq. (\ref{betaseries}) are independent of the scheme used for
regularization and renormalization, while the $b_n$ with $n \ge 3$ are
scheme-dependent \cite{bgz74,gross75}.  The coefficients $b_1$ and
$b_2$ for the one-component $\phi^3_6$ theory were calculated in
\cite{macfarlane_woo}, while $b_3$ was calculated in \cite{alcantara}
(in the $\overline{\rm MS}$ scheme \cite{msbar}).  It was observed
early on \cite{macfarlane_woo} that the one-component $\phi^3_6$
theory is asymptotically free, i.e., $g(\mu) \to 0$ in the ultraviolet
(UV) limit, $\mu \to \infty$.  For this theory, and
for a $\phi^3_6$ theory with a general global symmetry group $G$, the
beta function was calculated up to four-loop order, inclusive, in
Ref. \cite{gracey2015} \cite{bcn}. Recently, Ref.  \cite{gracey2020}
presented a four-loop calculation of the beta function for a
$\phi^3_6$ theory with a scalar transforming as a bi-adjoint
representation of a direct product $G_1 \otimes G_2$ global symmetry
group (see also \cite{onom}).

An important question is whether, for the region of $\bar g$ where a
perturbative calculation of the beta function is reliable, the beta
function of this theory exhibits evidence for a zero away from the
origin, at a physical, positive, value of $a$.  If the theory is
asymptotically free, this would be an infrared (IR) fixed point of the
renormalization group (IRFP), denoted $a_{_{IR}}$, while if the theory
is infrared free, this would be an UV fixed point of the RG
(UVFP), denoted $a_{_{UV}}$. In the UV-free case, one thus
considers the RG evolution of the theory from the deep UV.  If the
theory exhibits an IRFP, then as the reference momentum scale $\mu$
decreases from large values, $a=a(\mu)$ increases and approaches
$a_{_{IR}}$ from below as $\mu \to 0$. In the IR-free case, one
envisages starting the RG evolution from the IR; if the theory
exhibits an UVFP, then as $\mu \to \infty$, $a(\mu)$ approaches
$a_{_{UV}}$ from below.

In this paper we carry out an analysis of zeros of the respective beta
functions for three types of $\phi^3$ theories in $d=6$ spacetime
dimensions, namely those with a real, one-component scalar, a scalar
transforming according to the fundamental representation of SU($N$),
and a scalar transforming according according to a bi-adjoint
representation of ${\rm SU}(N) \otimes {\rm SU}(N)$.  The organization
of the paper is as follows. In Section \ref{methods_section} we
discuss some relevant methodology.  In Sections
\ref{1component_section}-\ref{biadjoint_section} we present our
results for the three $\phi^3_6$ theories under consideration. Our
conclusions are given in Section \ref{conclusion_section}.


\section{Methodology}
\label{methods_section}

In this section we briefly discuss some methodology that is relevant
for our study of the beta functions and their zeros in $\phi^3_6$
theories.  One carries out this study using the beta function
calculated (perturbatively) to a given finite $n$-loop order. The
maximal loop order to which one can carry out this study in a
scheme-independent manner is the two-loop order.  Thus, if $b_1$ and
$b_2$ are nonzero, then a necessary condition for a theory to exibit a
physical zero of the beta function at a nonzero value of $a$ is that
$b_1$ and $b_2$ must have opposite signs.  If the $n$-loop beta
function has more than one zero on the positive real $a$ axis, we
denote the one nearest to the origin as $a_{_{IR,n\ell}}$ or
$a_{_{UV,n\ell}}$ in the two respective cases of an asymptotically
free or infrared free theory. An additional necessary condition for
the $n$-loop beta function to exhibit robust evidence for an IR or UV
zero at the respective values $a_{_{IR,n\ell}}$ and $a_{_{UV,n\ell}}$
is that the beta functions calculated to $(n+1)$-loop order should
also exhibit a zero, and the fractional difference between the
$n$-loop and $(n+1)$-loop values should be small.

Before proceeding, for perspective, it is useful to mention two
examples where these conditions are satisfied.  The first example is a
non-Abelian gauge theory in $d=4$ spacetime dimensions with gauge
group $G$ containing $N_f$ massless fermions transforming according to
a given representation $R$ of $G$.  This theory is asymptotically free
for $N_f$ less than an upper $(u)$ bound, $N_u=11C_A/(4T_f)$
\cite{b1_nagt}, where $C_A=C_2(Adj)$ is the quadratic Casimir
invariant for the adjoint representation and $T_f=T(R)$ is the trace
invariant \cite{group_invariants,nintegral}.  There is a range of
values of $N_f$ less than $N_u$ where the two-loop beta function of
this theory has an IR zero \cite{b2_nagt,bz} at a value
$\alpha_{IR,2\ell}$ that goes to zero as $N_f$ (formally generalized
to nonzero real values) approaches $N_u$ from below.  For $N_f$ less
than, but close to $N_u$, the IR theory is weakly coupled, and one
expects that the values of this IR zero and of physical quantities
such as anomalous dimensions of gauge-invariant operators calculated
to finite order at the IRFP are reasonably stable with respect to the
inclusion of higher-loop terms in the beta function.  This has been
shown explicitly and quantitatively up to the four-loop \cite{bvh,ps}
and five-loop order \cite{flir,dexl}.  As $N_f$ decreases below the
region near $N_u$, the IR theory becomes more strongly coupled and
higher-order terms in perturbative series expansions become more
important.  Although the value of the IR zero, $\alpha_{IR,n\ell}$, at
$n$-loop order is scheme-dependent if $n \ge 3$, scheme-independent
series expansions as power series in the variable $\Delta_f = N_u-N_f$
have been used to obtain scheme-independent calculations of physical
quantities such as anomalous dimensions \cite{gtr,dexl,pgb,bgs}.  The
resultant values of these anomalous dimensions have been compared with
lattice simulations \cite{dexl,pgb}. (For reviews of lattice
measurements, see, e.g., \cite{lgtrev,simons}.)  This stability of
physical results for $N_f$ slightly below $N_u$ in such non-Abelian
gauge theories is the sort of necessary behavior that one would
require to certify the existence of an IRFP of the renormalization
group in an asymptoticaly free $\phi^3_6$ theory.

An example of a reliably calculated UVFP of the renormalization group
in an IR-free theory is provided by an exact solution of the O($N$)
nonlinear $\sigma$ model in the $N \to \infty$ limit, in
$d=2+\epsilon$ dimensions \cite{nlsm}, where $\epsilon$ is small.
This UVFP was calculated nonperturbatively by means of a summation of
an infinite number of Feynman diagrams in this $N \to \infty$ limit.

As noted above, for our analysis of the beta functions of the various
$\phi^3_6$ theories, we shall use the $b_n$ coefficients with $1 \le n
\le 4$, with $b_3$ and $b_4$ calculated in the $\overline{\rm MS}$
scheme \cite{msbar}, from
Refs. \cite{alcantara,gracey2015,gracey2020}.  Effects of scheme
transformations on beta function coefficients were calculated in
\cite{sch,sch23} (see also \cite{brodsky,tsch,graceysimms}.)


\section{One-Component $\phi$ Field}
\label{1component_section}

In this section we consider the $\phi^3_6$ theory with a one-component
(real) scalar field, with the Lagrangian density (\ref{lag1}). The
one-loop and two-loop coefficients in $\beta_{\bar g}$ are
\cite{macfarlane_woo}
\beq
b_1 = -\frac{3}{4} 
\label{b1_1component}
\eeq
and
\beq
b_2 = -\frac{5^3}{2^4 \cdot 3^2} = -0.8680556\ .
\label{b2_1component}
\eeq
The fact that $b_1$ is negative means that this theory is asymptotically free.
Since these coefficients have the same sign, the beta function has no IR zero
at the maximal scheme-independent level, namely the two-loop level. 

In the $\overline{\rm MS}$ scheme, the three-loop coefficient is
\cite{alcantara}
\beq
b_3 = -\frac{5}{2^3} \Big ( \frac{6617}{2^5 \cdot 3^4} + \zeta_3 \Big ) = -2.3468199 \ .
\label{b3_1component}
\eeq
In Eq. (\ref{b3_1component}) and similar equations we show the simple
factorizations of denominators. Although the numerator of $b_2$ happens to
have a simple factorization, most numerator numbers do not;
for example, $6617=13 \cdot 509$.  The
four-loop coefficient, again in the $\overline{\rm MS}$ scheme, is
\cite{gracey2015,bcn}
\beqs
b_4 &=& \frac{3404365}{2^{10} \cdot 3^6} + \frac{4891\zeta_3}{2^5 \cdot 3^3}
- \frac{15\zeta_4}{2^5} -\frac{5\zeta_5}{3} \cr\cr
&=& 9.129607 \ , 
\label{b4_1component}
\eeqs
where $\zeta_s = \sum_{n=1}^\infty n^{-s}$ is the Riemann zeta
function. (Here we could substitute $\zeta_4= \pi^4/90$, but we leave
the $\zeta_4$ term in its abstract form.) Since the $b_n$ with
$n \ge 3$ are scheme-dependent, so are the zeros of the $n$-loop
beta function for $n \ge 3$.  Nevertheless, one may check the zeros
of the three-loop beta function away from the origin. These are the
solutions of the quadratic equation $b_1+b_2 a + b_3 a^2=0$. We find these
solutions are a complex-conjugate pair and hence are unphysical. This
analysis at the two-loop and three-loop level provides strong evidence
against the existence of an IR zero in the beta function.  At the
four-loop level, the zeros of the beta function away from the origin,
which are the solutions to the cubic equation $b_1+b_2a+b_3a^2+b_4a^3=0$,
are comprised of a complex-conjugate pair and the value
$a=0.622134$. Because the one real positive root was not present at
either the maximal scheme-independent two-loop level or at the
three-loop level, we do not consider it as robust evidence for an IR
zero of the beta function.


\section{SU($N$) Theory with Scalar in Fundamental Representation}
\label{sun_section}

In this section we investigate the beta function of the $\phi^3_6$ theory
where $\phi$ transforms as the fundamental representation of a (global)
SU($N$) symmetry group. As noted above, we consider $N$ in the range
$N \ge 3$ since $d_{ijk}$ vanishes for SU(2), resulting in a free theory.
The beta function has been calculated to four-loop
order for this theory in \cite{gracey2015} \cite{bcn}.  The two scheme-independent
coefficients are \
\beq
b_1 = -\frac{(N^2-20)}{4N}
\label{b1sun}
\eeq
and
\beq
b_2 = -\frac{(5N^4-496N^2+5360)}{2^4 \cdot 3^2 N^2} \ .
\label{b2sun}
\eeq
The one-loop coefficient $b_1$ is positive for small $N$ and passes through
zero to negative values as $N$ (formally generalized from integral values
$N \ge 2$ to positive real values \cite{nintegral}) increases through the value
\beq
N_{b1z} = 2\sqrt{5} = 4.4721360 \ ,
\label{Nb1z}
\eeq
where this and other floating-point numbers are quoted to the
indicated accuracy, and the subscript $b1z$ stands for ``$b_1$ zero''.
Thus, this theory is IR-free for $N < N_{b1z}$ and UV-free (i.e.,
asymptotically free) for $N > N_{b1z}$.  The physical reason for the
change in the sign of $b_1$ and the resultant change in the
renormalization-group behavior as $N$
increases through the value $2\sqrt{5}$ can be traced to the
individual contributions to $b_1$ from one-loop two-point and
three-point Feynman diagrams.  This is evident from the expression for
$b_1$ in terms of group invariants, namely \cite{gracey2020}
\beq
b_1 = \frac{1}{4}(T_2 - 4T_3) \ , 
\label{b1form}
\eeq
where $T_2$ and $T_3$ are defined by the traces 
\beq
d^{i i_1 i_2} d^{j i_1 i_2} = T_2 \delta^{ij}
\label{t2def}
\eeq
and
\beq
d^{i i_1 i_2} d^{j i_1 i_3} d^{k i_2 i_3} = T_3 d^{ijk} \ .
\label{t3def}
\eeq
The traces $T_2$ and $T_3$ occur in the one-loop corrections to the
two-point and three-point functions. For SU($N$) \cite{macfarlane1968}
\beq
T_2 = \frac{N^2-4}{N}
\label{t2sun}
\eeq
and
\beq
T_3 = \frac{N^2-12}{2N} \ . 
\label{t3sun}
\eeq
The interplay of both of these types of corrections determines $b_1$.

The coefficient $b_2$ is negative for $N < N_{b2z,-}$, positive in the interval
$N_{b2z,-} < N < N_{b2z,+}$, and negative for $N > N_{b2z,+}$, where 
\beq
N_{b2z,\pm}=\frac{2}{\sqrt{5}} \sqrt{62 \pm 3\sqrt{241}} \ .
\label{Nb2z}
\eeq
Numerically,
\beq
N_{b2z,-}= 3.5131155, \quad N_{b2z,+}= 9.319765 \ .
\label{Nb2znum}
\eeq
Consequently, this theory thus has four different regimes of RG behavior, depending
on the value of $N$ (again, formally generalized to positive real values):

\begin{enumerate}

\item $3 \le N < 3.513$: IR-free with, a UV zero of the beta function
  $\beta_{\bar g,2\ell}$ 

   \item $3.513 < N < 4.472$: IR-free, with no UV zero of $\beta_{\bar g,2\ell}$ 

   \item $4.472 < N < 9.320$: UV-free, with an IR zero of $\beta_{\bar g,2\ell}$

   \item $N > 9.320$: UV-free, with no IR zero of $\beta_{\bar g,2\ell}$.

\end{enumerate}
These properties are summarized in Table \ref{sun_properties_table}.
The respective real intervals in $N$ contain the physical integer
values (i) $N=3$; (ii) $N=4$; (iii) $5 \le N \le 9$; and (iv) $N
\ge 10$.

We first consider the asymptotically free regime, defined by the inequality
$N > N_{b1z}$. For $N$ in the interval
$N_{b1z} < N < N_{b2z,+}$, i.e., $4.472 < N < 9.320$, 
the two-loop beta function $\beta_{\bar g,2\ell}$ has a IR zero at
$a=-b_1/b_2 \equiv a_{_{IR,2\ell}}$, where 
\beq
a_{_{IR,2\ell}} = \frac{36N(N^2-20)}{-5N^4+496N^2-5360} \ .
\label{air_2loop_sun}
\eeq
In Table \ref{air_nloop_sun_table} we list values of $a_{_{IR,2\ell}}$ for
integer values of $N$ in the interval $N_{b1z} < N < N_{b2z,+}$.
The calculation leading to this IR zero is expected to be most reliable
toward the lower end of this interval, where $a_{_{IR,2\ell}}$ is small,
and to become less reliable toward the upper end of the interval, where
$a_{_{IR,2\ell}}$ grows to larger values.

To investigate how stable this IR zero of the two-loop beta function
$\beta_{\bar g,2\ell}$ is to the inclusion of higher-order terms, we 
examine the three-loop and four-loop beta functions, $\beta_{\bar g,3\ell}$
and $\beta_{\bar g,4\ell}$.  For this purpose, we make
use of the expressions for $b_3$ and $b_4$, as calculated in the
$\overline{\rm MS}$ scheme, from Ref. \cite{gracey2015} \cite{,bcn},
\begin{widetext}
\beqs
b_3 = \frac{1}{2^8 \cdot 3^4 N^3}\Big [ -211N^6 + (27132-62208\zeta_3)N^4
  +(-1220688+20736\zeta_3)N^2 + 9272896+4396032\zeta_3 \Big ] 
\label{b3sun}
\eeqs
and
\beqs
b_4 &=& \frac{1}{2^{10} \cdot 3^6 N^4} \Big [ (327893+870048\zeta_3 -
  1321920\zeta_5)N^8 + (-8142840 - 14427072\zeta_3 -559872\zeta_4+
  31570560\zeta_5)N^6 \cr\cr
  &+& (112740480+155416320\zeta_3+11384064\zeta_4-421770240\zeta_5)N^4 \cr\cr
  &+& (-1264882304-1477343232\zeta_3+35831808\zeta_4+1950842880\zeta_5)N^2
  \cr\cr
  &+& 5761837824+7029669888\zeta_3-791285760\zeta_4+995328000\zeta_5 \Big ] \ .
\label{b4sun}
\eeqs
\end{widetext}
As before, the zeros of $\beta_{\bar g,3\ell}$ away from the origin
are the solutions of the equation $b_1+b_2a+b_3a^2=0$. In the interval
$N_{b1z} < N < N_{b2z,+}$ under consideration here, the discriminant
$b_2^2-4b_1b_3$ (which is a quartic function in the variable $N^2$) is
positive for a very small interval $N_{b1z} < N < N_{dz}$, but passes
through zero to negative values as $N$ increases through the value
\beq
N_{dz} = 4.497050
\label{Ndz}
\eeq
(where the subscript $dz$ stands for ``discriminant zero''). 
Hence, except for the very small
interval $4.472 < N < 4.497$, the zeros of $\beta_{\bar g,3\ell}$
(away from the origin) are comprised of a complex-conjugate pair of
$a$ values and are thus unphysical. This is indicated in Table
\ref{air_nloop_sun_table}.  These results show that although the
two-loop beta function exhibits an IR zero in this interval $4.472 < N
< 9.320$, it is not stable to the inclusion of higher-order
perturbative corrections.  We also calculate the zeros of the
four-loop beta function, given as the roots of the equation $b_1 +
b_2a + b_3 a^2 + b_4 a^3=0$.  The results are listed in Table
\ref{air_nloop_sun_table}. As is evident, they consist of a real value
and an unphysical complex-conjugate pair of roots.  The fact that the
real value is not at all close to $a_{_{IR,2\ell}}$ provides further
evidence against a robust IR zero of the beta function.

In the interval $N > N_{b2z,+}$, i.e., $N>9.320$, this SU($N$)
$\phi^3_6$ theory does not have an IR zero at the maximal
scheme-independent level of two loops.

Finally, we consider the interval $N < N_{b1z}$, i.e., $3 \le N <
4.472$, where the theory is IR-free.  As discussed above, in the
sub-interval $N_{b2z,-} < N < N_{b1z}$, i.e., $3.513 < N < 4.472$,
there is no UV zero in the two-loop beta function, $\beta_{\bar
  g,2\ell}$.  In the sub-interval $3 \le N < 3.513$, including the
physical integral value $N=3$, $\beta_{\bar g,2\ell}$ does have a UV
zero, denoted $a_{_{UV,2\ell}}$, which is given by the right-hand side
of Eq. (\ref{air_2loop_sun}). This two-loop zero, $a_{_{UV,2\ell}}$,
has the value $1188/1301=0.913144$ for $N=3$.  However, in order for
this to be considered as a reliable UV zero of the beta function, it
is necessary that the value should be reasonably stable when one
includes higher-order terms in the beta function.  We find that this
is not the case. At the three-loop level, the zeros of $\beta_{\bar
  g,3\ell}$ away from the origin for this $N=3$ case consist of an
unphysical complex-conjugate pair of values of $a$. At the four-loop
level, the three roots of the equation $b_1+b_2a+b_3a^2+b_4a^3=0$ are
comprised of a negative value and a complex-conjugate pair, all of
which are unphysical.  We list these results in Table
\ref{air_nloop_sun_table}.  Consequently, although the beta function
of this SU($3$) $\phi^3_6$ theory does exhibit a UV zero at the
two-loop level, it does not satisfy the requirement of being stable to
higher-loop corrections.

The large-$N$ limit of this theory is also of interest.  We define
the rescaled coupling
\beq
\xi \equiv \bar g N
\label{xi}
\eeq
and the corresponding beta function
\beq
\beta_\xi = \frac{d\xi}{d\ln\mu} \ .
\label{betaxi}
\eeq
This beta function has the series expansion
\beq
\beta_\xi = \xi \sum_{n=1}^\infty \hat b_n \hat a^n
\label{betaxiseries}
\eeq
where
\beq
\hat b_n = \lim_{N \to \infty} \frac{b_n}{N^n}
\label{bellhat}
\eeq
and
\beq
\hat a = \xi^2 \ .
\label{ahat}
\eeq
From the expressions for the $b_n$, $1 \le n \le 4$ we have
\beq
\hat b_1 = -\frac{1}{4}
\label{b1hat}
\eeq
\beq
\hat b_2 = -\frac{5}{2^4 \cdot 3^2} = -(3.472222 \times 10^{-2})
\label{b2hat}
\eeq
\beq
\hat b_3 = -\frac{211}{2^8 \cdot 3^4} = -(1.01755 \times 10^{-2})
\label{b3hat}
\eeq
and
\beqs
\hat b_4 &=& \frac{1}{2^4 \cdot 3}\Big ( \frac{327893}{2^6 \cdot 3^5} +
\frac{1007\zeta_3}{2 \cdot 3^2}-85\zeta_5 \Big ) \cr\cr
&=& 4.02503 \times 10^{-3} \ .
\label{b4hat}
\eeqs
In the large-$N$ limit, this theory has no IRFP at the
maximal, 2-loop scheme-independent level, since $\hat b_1$ and $\hat
b_2$ have the same sign.


\section{${\rm SU}(N) \otimes {\rm SU}(N)$ Theory with Bi-adjoint Scalar} 
\label{biadjoint_section}

The condition that a theory is UV-free or IR-free is that for small
(physical) values of the coupling near the origin, its beta function
is negative or positive, respectively.  If, as is usually the case,
the one-loop coefficient, $b_1$, is nonzero, this is equivalent to the
condition that $b_1$ is negative or positive, respectively.
In a theory where $b_1$ depends on a parameter, such as the SU($N$)
$\phi^3_6$ theory, one may formally choose this parameter so that $b_1=0$
and then examine the sign of $b_2$.  For example, in the case of the
SU($N$) $\phi^3_6$ theory, if one formally generalizes $N$ from the
physical range of integers $N \ge 3$ to non-negative real numbers and
sets $N=N_{b1z}=2\sqrt{5}$, this renders $b_1=0$ in Eq. (\ref{b1sun}).
Substituting this value of $N$ into $b_2$, one obtains $b_2=8/9 > 0$,
so that the theory is IR-free.  Of course, this is just a formal result,
since it depends on setting $N$ to a non-integer value.

Recently, Ref. \cite{gracey2020} reported a physical example of a
theory with an identically zero one-loop term, i.e., $b_1=0$.  This is
a $\phi^3_6$ theory with a bi-adjoint (BA) scalar, i.e., a scalar
transforming according to the representation $(Adj,Adj)$ of a
direct-product global symmetry group $G_1 \otimes G_2$.  For our
purposes, it will suffice to consider the diagonal case where the
symmetry group is $G \otimes G$ and, furthermore, to take $G={\rm
  SU}(N)$, with $N \ge 2$.  We denote the scalar as $\phi^{a_1 a_2}$,
where here $1 \le a_1, \ a_2 \le N^2-1$. The Lagrangian density for
this theory is
\beqs
    {\cal L}_3 &=&
    \frac{1}{2}(\partial_\mu \phi^{a_1 a_2})(\partial^\mu \phi^{a_1a_2}) \cr\cr
    &-& \frac{g}{3!} f^{a_1 b_1 c_1} f^{a_2 b_2 c_2}
    \phi^{a_1 a_2} \phi^{b_1 b_2}\phi^{c_1 c_2} \ , 
    \label{lag3}
    \eeqs
where $f^{abc}$ are the structure constants of the Lie algebra of
${\rm SU}(N)$ \cite{group_invariants}.

The two-loop coefficient of the beta
function in this theory is \cite{gracey2020}
\beq
b_2^{(BA)} = -\frac{5N^4}{2^6 \cdot 3^2} \ .
\label{b2adj}
\eeq
Thus, this theory is asymptotically free.  For its study of this
theory, Ref. \cite{gracey2020} also calculated the three-loop and
four-loop terms in the beta function (in the $\overline{\rm MS}$
scheme).  These are
\beq
b_3^{(BA)} = \frac{N^2}{2^8 \cdot 3^4}\Big ( 55N^4 + 7776N^2 + 139968\zeta_3 \Big )
\label{b3_adj}
\eeq
and
\begin{widetext}
\beqs
b_4^{(BA)} &=& \frac{N^4}{2^{14} \cdot 3^6} \, \Big [ (-298081-825120\zeta_3+1244160\zeta_5)N^4
  + (-7091712-22394880\zeta_3+33592320\zeta_5)N^2 \cr\cr
  &-& 214990848\zeta_3+268738560\zeta_5 \ \Big ] \ .
\label{b4_adj}
\eeqs
\end{widetext}
Because the $b_n^{(BA)}$ are scheme-dependent for $n \ge 3$, it is not
possible to give a scheme-independent answer to the question of
whether the (perturbatively computed) beta function has an IR zero
in this theory.  With $b_3$ calculated in the $\overline{\rm MS}$
scheme, the theory has an IR zero in the 3-loop beta function, at
$a=-b_3^{(BA)}/b_2^{(BA)}$, i.e.,
\beq
a_{_{IR,3\ell,BA}} = \frac{180N^2}{55N^4 + 7776N^2 + 139968} \ .
\label{air_3loop_adj}
\eeq
We list values of $a_{_{IR,3\ell,BA}}$ in Table
\ref{air_nloop_biadjoint_table} for an illustrative range of values of
$N$.

We now investigate the effect of including the next-higher-order term,
namely, the four-loop term, in the beta function.  The condition that the
four-loop beta function vanishes for $a$ away from the origin is the
equation $b_2^{(BA)}+b_3^{(BA)}a+b_4^{(BA)}a^2=0$. Using the expressions for 
$b_3^{(BA)}$ and $b_4^{(BA)}$ in Eqs. (\ref{b3_adj}) and (\ref{b4_adj}), we
find that one solution for $a_{_{IR,4\ell,BA}}$ is quite close to the three-loop
value, $a_{_{IR,3\ell,BA}}$.  We list this solution in Table \ref{air_nloop_biadjoint_table}
together with the corresponding fractional difference $\delta_{IR,3,4}$, where
\beq 
\delta_{IR,n,n+1} \equiv \frac{a_{_{IR,(n+1)\ell,BA}}-a_{_{IR,n\ell,BA}}}{a_{_{IR,n\ell,BA}}} \ .
\label{delta}
\eeq
This agreement of the three-loop and four-loop values of the IR zero of beta in
the $\overline{\rm MS}$ scheme was noted for $N=3$ in \cite{gracey2020}, and
here it is extended to other values of $N$.

We may also consider the large-$N$ limit of this theory.  For this purpose we
define the variable
\beq
\eta \equiv \bar g N^2
\label{eta}
\eeq
and the corresponding beta function
\beq
\beta_\eta = \frac{d\eta}{d\ln\mu} \ .
\label{betaeta}
\eeq
This has the series expansion
\beq
\beta_\eta = \eta \sum_{n=1}^\infty \tilde b^{(BA)}_n \eta^n \ , 
\label{betaeta_series}
\eeq
where
\beq
\tilde b^{(BA)}_n = \lim_{N \to \infty} \frac{b^{(BA)}_n}{N^{2n}} \ . 
\label{badjhat}
\eeq
We have $\tilde b_1^{(BA)}=0$, and
\beq
\tilde b_2^{(BA)} = -\frac{5}{2^6 \cdot 3^2} = 0.868056 \times 10^{-2}
\label{b2hat_adj}
\eeq
\beq
\tilde b_3^{(BA)} = \frac{55}{2^8 \cdot 3^4} = 2.65239 \times 10^{-3}
\label{b3hat_adj}
\eeq
and
\beqs
\tilde b_4^{(BA)} &=& \frac{1}{2^{14} \cdot 3^6} \Big ( -298081 - 825120\zeta_3
+1244160\zeta_5 \Big ) \cr\cr
&=& 1.52248 \times 10^{-5} \ .
\label{b4hat_adj}
\eeqs
The IR zeros of the rescaled three-loop and four-loop beta functions are,
respectively, 
\beq
\eta_{_{IR,3\ell,BA}} = \frac{36}{11} = 3.272727
\label{eta_3loop_adj}
\eeq
and
\begin{widetext}
\beq
\eta_{_{IR,4\ell,BA}} = \frac{144 \Big [ -110 + \Big (5(-295661-825120\zeta_3 + 1244160\zeta_5)\Big )^{1/2} \Big ] }
    {-298081-825120\zeta_3+1244160\zeta_5} = 3.2134506 \ .
\label{xir_4loop_adj}
\eeq
\end{widetext}
The fractional difference between these is reasonably small:
\beq
\frac{\eta_{_{IR,4\ell,BA}}-\eta_{_{IR,3\ell,BA}}}{\eta_{_{IR,3\ell,BA}}} =
-(1.8112 \times 10^{-2}) \ .
\label{delta_34_adj}
\eeq
This small fractional difference can be understood as a consequence of
the fact that $\tilde b_4^{(BA)}$ is much smaller than $\tilde
b_3^{(BA)}$.  These results are consistent with the inference that in
the $N \to \infty$ limit, this theory has an IR zero in the beta
function. However, one must treat this inference with considerable
caution, since it involves scheme-dependent beta function terms in an
essential way.


\section{Conclusions} 
\label{conclusion_section}

In this work we have investigated whether the beta functions for three
$\phi^3_6$ theories exhibit robust evidence for zeros away from the
origin.  The one-component theory is asymptotically free, and has an
IR zero at the two-loop level.  However, we find that it is not stable
to the inclusion of three-loop and four-loop terms in the beta
function, and hence we conclude that there is not persuasive evidence
for a robust IR zero in this theory.  For the $\phi^3_6$ theory with a
scalar transforming according to the fundamental representation of a
(global) SU($N$) symmetry group with $N \ge 3$, we find four different
types of renormalization-group behavior, depending on the value of
$N$.  In particular, for $N$ (generalized from positive integers to
positive real numbers) in the interval $4.47 < N < 9.32$, the theory
is asymptotically free and has an IR zero in the two-loop beta
function, but we find that it is not stable to the inclusion of
higher-loop terms.  For $N$ in the interval $3 \le N < 3.51$, the
theory is IR-free and has a UV zero in the two-loop beta function, but
we again find that this is not stable to the inclusion of higher-loop
terms.  In the two other intervals, namely $3.51 < N < 4.47$ and $N >
9.32$, the one-loop and two-loop terms in the beta function have the
same sign, so the beta function has no physical zero away from the
origin in coupling-constant space.  The third $\phi^3_6$ theory that
we consider features a scalar transforming as a bi-adjoint
representation of a global ${\rm SU}(N) \otimes {\rm SU}(N)$ symmetry
with $N \ge 2$. This theory has the property that the one-loop term in
the beta function vanishes and the two-loop term is negative, so the
theory is asymptotically free.  For this theory, the question of
whether the higher-loop beta function has an IR zero cannot be
answered in a scheme-independent way, and hence results must be
treated with the requisite caution.  Nevertheless, we do find
that in the $\overline{\rm MS}$ scheme, the three-loop and four-loop calculations
yield values of an IR zero in reasonable agreement with each other.


\begin{acknowledgments}

  This research was supported in part by a DFG
  (Deutsche Forschungsgemeinschaft)
  Fellowship (J.A.G.) and by the U.S. NSF Grant NSF-PHY-1915093 (R.S.).

\end{acknowledgments}


\bigskip
\bigskip
\bigskip

\newpage

\begin{table}
  \caption{ \footnotesize{Regimes of different behavior of $\phi^3_6$ theory
      with SU($N$) global symmetry, as a function of $N$ (formally generalized from
      positive integer to positive real values \cite{nintegral}).}}
\begin{center}
\begin{tabular}{|c|c|c|c|} \hline\hline
  $N$ & $b_1$ & $b_2$ & properties \\
  \hline
$3 \le N_{b2z,-}$, i.e.  & $+$ & $-$ & IR-free with $a_{_{UV,2\ell}}$ \\
  $3 \le N < 3.513$      &     &     &                             \\
  \hline
$N_{b2z,-}<N<N_{b1z}$, i.e.& $+$ & $+$& IR-free, no $a_{_{UV,2\ell}}$ \\
  $3.513 < N < 4.472$  &      &    &                               \\
  \hline
$N_{b1z}<N<N_{b2z,+}$, i.e. & $-$ & $+$ & UV-free with $a_{_{IR,2\ell}}$ \\
  $4.472<N<9.320$      &      &    &                               \\
  \hline
$N>N_{b2z,+}$, i.e.   & $-$ & $-$  & UV-free, no $a_{_{IR,2\ell}}$ \\
$N > 9.320$            &      &    &                              \\  
\hline\hline
\end{tabular}
\end{center}
\label{sun_properties_table}
\end{table}
%

\begin{table}
\caption{\footnotesize{Values of zeros of the $n$-loop
    beta function, $\beta_{\bar g,n\ell}$ away from the origin, in the
    variable $a=(\bar g)^2$, for $2 \le n \le 4$, as a function of $N$,
    in the SU($N$) $\phi^3_6$
    theory with a scalar field transforming as the fundamental representation of
    SU($N$). The notation ``u'' denotes unphysical zeros (which are comprised of
complex-conjugate pairs here). If $N=4$ or $N \ge 10$, the theory has no
scheme-independent zero of the beta function, and hence these cases are
not tabulated. See text for further details.}}
\begin{center}
\begin{tabular}{|c|c|c|c|} \hline\hline
$N$ & 2-loop & 3-loop & 4-loop \\
  \hline
3  &  0.913  &  u  & u, \ $-0.676$  \\
\hline
5  &  0.230   &  u   & u, \ 0.495 \\
6  &  0.574   &  u   & u, \ 0.224 \\
7  &  1.053   &  u   & u, \ 0.190 \\
8  &  2.146   &  u   & u, \ 0.175 \\
9  &  9.828   &  u   & u, \ 0.166 \\
\hline\hline
\end{tabular}
\end{center}
\label{air_nloop_sun_table}
\end{table}

\begin{table}
\caption{\footnotesize{Values of IR zeros $a_{_{IR,n\ell,BA}}$ of the $n$-loop
 beta function, $\beta_{\bar g,n\ell}$ away from the origin, in the
 variable $a=(\bar g)^2$, for $2 \le n \le 4$, in the
 as a function of $N$, in the ${\rm SU}(N) \otimes {\rm SU}(N)$ $\phi^3_6$
 theory with a scalar field transforming as a bi-adjoint representation. 
The notation ``u'' denotes an unphysical zeros (which are comprised of
complex-conjugate pairs here). The notation $x$e-2 means $x \times 10^{-2}$.
See text for further details.}}
\begin{center}
\begin{tabular}{|c|c|c|c|} \hline\hline
$N$ & $a_{_{IR,3\ell,BA}}$ & $a_{_{IR,4\ell,BA}}$ & $\delta_{IR,3,4}$ \\
\hline
2  & 3.596e-3   & 3.585e-3  &  $-2.916$e-3 \\
3  & 6.675e-3   & 6.598e-3  &  $-1.160$e-2 \\
4  & 0.9389e-2  & 0.9136e-2 &  $-2.690$e-2 \\
5  & 1.133e-2   & 1.081e-2  &  $-4.609$e-2 \\
6  & 1.247e-2   & 1.166e-2  &  $-6.556$e-2 \\
7  & 1.295e-2   & 1.187e-2  &  $-8.281$e-2 \\
8  & 1.293e-2   & 1.168e-2  &  $-9.670$e-2 \\
9  & 1.258e-2   & 1.112e-2  &  $-0.1070$   \\
10 & 1.203e-2   & 1.066e-2  &  $-0.1142$   \\
\hline\hline
\end{tabular}
\end{center}
\label{air_nloop_biadjoint_table}
\end{table}



\begin{thebibliography}{99}
\bibitem{bcn}
  To agree with the standard definition, the signs of the
  beta functions coefficients
  given in \cite{gracey2015} should be 
  modified as follows: $b_n \to -b_n$ for $n=1,3,4$, with the sign of
  $b_2$ unchanged. With our
  definition (\ref{beta}), the beta function coefficients in
  \cite{gracey2015} and \cite{gracey2020} are also multiplied by 2.
\bibitem{group_invariants}
%
We recall the general definitions of these group invariants. Denote
$T_R$ as a generator of the Lie algebra of a group $G$ in the
representation $R$ with dimension $d_R$. Then the quadratic Casimir
invariant $C_2(R)$ is defined by $T^a_RT^a_R = C_2(R) I$, where $I$ is
the $d_R \times d_R$ identity matrix, and the trace invariant $T(R)$
is defined by ${\rm Tr}_R(T^a_R T^b_R) = T(R)\delta^{ab}$, where $1
\le a \le o(G)$, with $o(G)$ the order of the group.  We write
$C_A=C_2(Adj)$ and $C_f = C_2(R)$.  For SU($N$),
$C_2(Adj)=T(Adj)=N$. Denoting the generator of the Lie algebra of
SU($N$) in the fundamental representation as $T^a$, the structure
constant $f^{abc}$ is defined by $[T^a,T^b]=if^{abc}T^c$ and the
symmetric tensor $d^{abc}$ is defined by $d^{abc}={\rm Tr}(T^a
\{T^b,T^c\})$, with the normalization ${\rm Tr}(T^aT^b)=
(1/2)\delta^{ab}$.
\bibitem{nintegral}
%
Here and below, when expressions are given for $N$ that evaluate
to non-integral real values, it is understood that they are formal and are
interpreted via an analytic continuation of $N$ from physical nonnegative
integer values to real numbers.
\end{thebibliography}
\end{document}